\documentclass{iaus}
\usepackage{txfonts}
\usepackage{graphicx}

\title[Massive galaxies with very young AGN] 
{Massive galaxies with very young AGN}

\author[N. de Vries et al.]   
{Nathan de Vries$^1$, I.A.G. Snellen$^1$, R.T. Schilizzi$^{1,2}$, \break
  M.D. Lehnert$^3$ \and M.N. Bremer$^4$}

\affiliation{$^1$Leiden Observatory, Leiden University, PO Box 9513, 2300 RA, Leiden, The Netherlands \break email: vriesn@strw.leidenuniv.nl\\[\affilskip]
$^2$International SKA Project office, PO Box 2, 7990 AA, Dwingeloo, The Netherlands\\[\affilskip]
$^3$GEPI, Observatoire de Paris, 5 place Jules Janssen, 92195 Meudon, France\\[\affilskip]
$^4$Department of Physics, Bristol University, Tyndall Avenue, Bristol BS8 1TL, UK
}

\pubyear{2007}
\volume{245}  
\pagerange{119--126}
\date{?? and in revised form ??}
\jname{Formation and Evolution of Galaxy Bulges}
\editors{M. Bureau, E. Athanassoula \& B. Barbuy, eds.}
\begin{document}

\maketitle

\begin{abstract}
Gigahertz Peaked Spectrum (GPS) radio galaxies are generally thought to be the young counterparts of classical extended radio sources and live in massive ellipticals. GPS sources are vital for studying the early evolution of radio-loud AGN, the trigger of their nuclear activity, and the importance of feedback in galaxy evolution. We study the ‘Parkes half-Jansky’ sample of GPS radio galaxies of which now all host galaxies have been identified and 80\% has their redshifts determined (0.122 $<$ z $<$ 1.539). Analysis of the absolute magnitudes of the GPS host galaxies show that at z $>$ 1 they are on average a magnitude fainter than classical 3C radio galaxies. This suggests that the AGN in young radio galaxies have not yet much influenced the overall properties of the host galaxy. However their restframe UV luminosities indicate that there is a low level of excess as compared to passive evolution models.
\end{abstract}

\firstsection 
\section{GPS host galaxies are among the most massive ellipticals}
GPS sources are very compact ($<1''$) radio sources with morphologies similar to extended Fanaroff \& Riley I/II radio sources (FRI/IIs). VLBI monitoring of GPS sources has shown these sources to expand, implying source ages of $10^{2-3}$ years only, and likely to grow out to become FRI/IIs. This makes GPS sources key objects for studying the early evolution of radio-loud AGN, the trigger of their nuclear activity, and the importance of feedback in galaxy evolution.

We investigate the Parkes Half-Jansky sample, consisting of 48 GPS galaxies (Snellen et al. 2002).
All host galaxies have now been identified and 39 have their redshift determined, which are in the range 0.122 $<$ z $<$ 1.539 (\cite{deVries07}). Fig. 1 shows the Hubble diagram of this sample (solid squares). The new data at z$\sim$1 confirm that GPS galaxies are on average 1.0 mag fainter in this
redshift range than 3C radio galaxies (open circles). This agrees with the hypothesis that GPS
galaxies are fainter and redder, due to the lack of the extra, blue light associated with the radio-optical
alignment effect. For comparison, a subset of the Luminous Red Galaxies (LRGs; small circles)
sample (Eisenstein et al. 2001) from the Sloan Digital Sky Survey (SDSS) are shown. The LRGs
form a volume limited sample of the most luminous, intrinsically red galaxies out to z=0.55 and
are thought to represent the most massive early type galaxies, many of which are classified as
Brightest Cluster Galaxies. This figure shows that host galaxies of GPS radio sources have similar
optical luminosities to LRGs, indicating that powerful young radio sources are hosted by the most
massive early type galaxies.

\begin{figure}[ht] 
\centering
 \includegraphics[height=7cm]{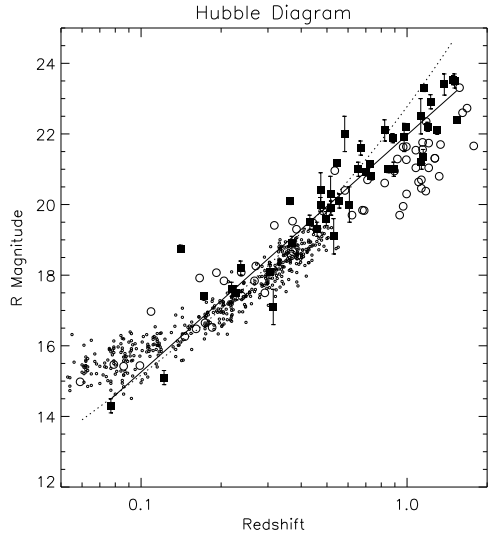}
\caption{Cousins $R_C$ Hubble diagram of GPS galaxies (solid squares),
3C galaxies (open circles), and LRGs (small circles).}
\end{figure}

\section{Recent starbursts in galaxies with young AGN?}

We compare the absolute magnitudes of GPS galaxies (solid
squares) and 3C radio galaxies (open circles) to passive evolution
models of single burst stellar populations. Fig. 2 (\textit{left}) shows
stellar population models with solar metallicity and different
formation redshifts ($z_f$=1.5, 2.0 and 3.0). Fig. 2 (\textit{right}) shows
models with different metallicities and a formation redshift of
$z_f$=1.5. Our data are in best agreement with a recent formation
redshift of 1.5-2.0, with a possible range of metallicities.
However, it would be surprising if the entire host galaxies would have formed so recently, especially because fig. 1 indicates that GPS galaxies are massive early type galaxies. Therefore we
interpret this result as evidence for recent starburst activity
and/or low level AGN induced light.

\begin{figure}[h] 
\centering
\hbox{
 \includegraphics[height=6.5cm]{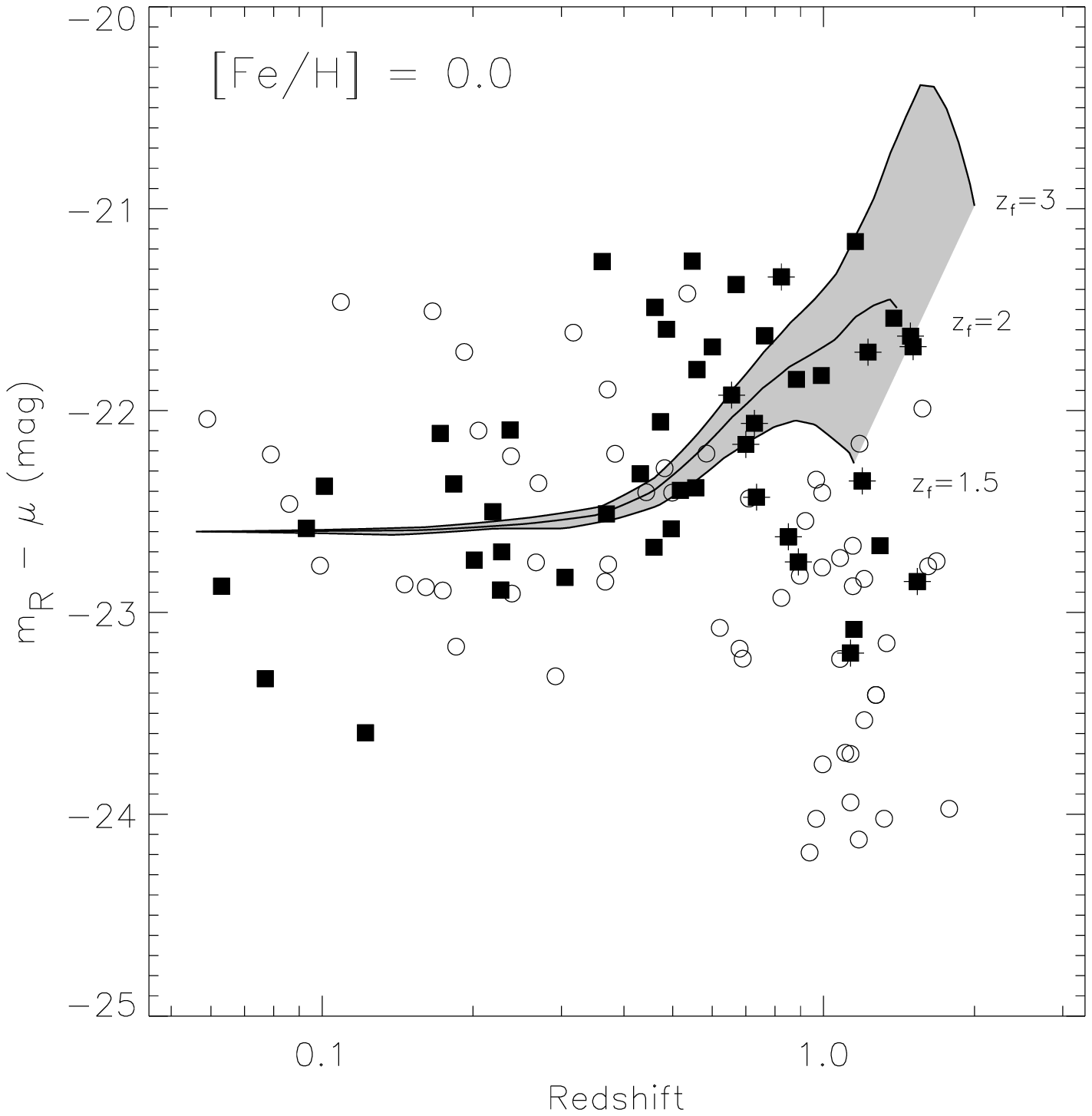}
 \includegraphics[height=6.5cm]{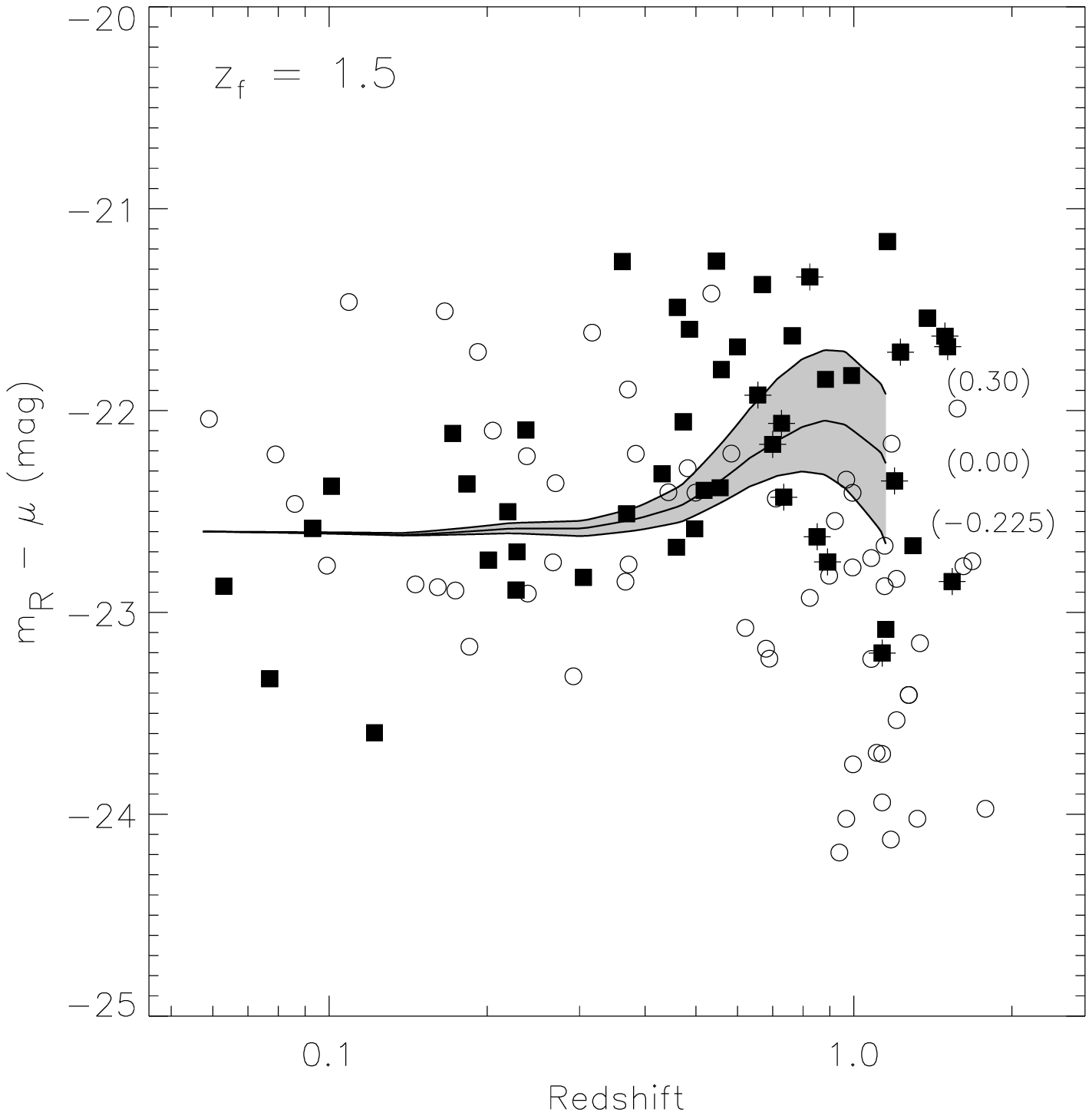}
}
\caption{Absolute magnitudes (without k-correction) of GPS galaxies (solid squares) and 3C galaxies (open
circles), compared with passive evolution models with different metallicities and formation redshifts.}
\end{figure}


\begin{thebibliography}{}
\bibitem[de Vries et al., 2007]{deVries07} de Vries N., Snellen I.A.G., Schilizzi R.T., Lehnert M.D., Bremer M.N., 2007, A\&A 464, 879
\bibitem[2001]{Eisenstein} Eisenstein D.J. et al., 2001, AJ, 122, 2267
\bibitem[2002]{Snellen02} Snellen I.A.G., Lehnert M.D., Bremer M.N., Schilizzi R.T., 2002,
           MNRAS, 337, 981
\end{thebibliography}
\end{document}